# 3D Reconstruction by Looking: Instantaneous Blind Spot Detector for Indoor SLAM through Mixed Reality

Hanbeom Chang[1,2], Jongseong Brad Choi[1,2*], Chul Min Yeum[3]

[1] Department of Mechanical Engineering, State University of New York, Stony Brook, United States
[2] Department of Mechanical Engineering, State University of New York, Korea, South Korea
[3] Department of Civil and Environmental Engineering, University of Waterloo, Canada

**Correspondence to:**
Jongseong Brad Choi
    119, Songdomunhwa-ro, Yeonsu-gu, Incheon, Republic of Korea (21985)
    Email: jongseong.choi@stonybrook.edu

**Author Contributions:** Conceptualization, Chul Min Yeum, and Jongseong Brad Choi; Methodology, Hanbeom Chang; Validation, Hanbeom Chang, Jongseong Brad Choi, and Chul Min Yeum; Writing and Paper Preparation, Hanbeom Chang, Chul Min Yeum, Jongseong Brad Choi; Supervision, Jongseong Brad Choi and Chul Min Yeum.

**Funding:** We would like to acknowledge support in part from the National Research Foundation of Korea under Grant No. 2022M1A3C2085237

## Abstract

Indoor SLAM often suffers from issues such as scene drifting, double walls, and blind spots, particularly in confined spaces with objects close to the sensors (e.g., LiDAR and cameras) in reconstruction tasks. Real-time visualization of point cloud registration during data collection may help mitigate these issues, but a significant limitation remains in the inability to in-depth compare the scanned data with actual physical environments. These challenges obstruct the quality of reconstruction products, frequently necessitating revisit and rescan efforts. For this regard, we developed the LiMRSF (LiDAR-MR-RGB Sensor Fusion) system, allowing users to perceive the in-situ point cloud registration by looking through a Mixed-Reality (MR) headset. This tailored framework visualizes point cloud meshes as holograms, seamlessly matching with the real-time scene on see-through glasses, and automatically highlights errors detected while they overlap. Such holographic elements are transmitted via a TCP server to an MR headset, where it is calibrated to align with the world coordinate, the physical location. This allows users to view the localized reconstruction product instantaneously, enabling them to quickly identify blind spots and errors, and take prompt action on-site. Our blind spot detector achieves an error detection precision with an F1 Score of 75.76% with acceptably high fidelity of monitoring through the LiMRSF system (highest SSIM of 0.5619, PSNR of 14.1004, and lowest MSE of 0.0389 in the five different sections of the simplified mesh model which users visualize through the LiMRSF device's see-through glasses). This method ensures the creation of detailed, high-quality datasets for 3D models, with potential applications in Building Information Modeling (BIM) but not limited.

## 1. Introduction

Simultaneous Localization and Mapping (SLAM) has been studied for decades. With significant advancements in computing power and the availability of low-cost vision sensors, applications in various fields have been found. Different variations of SLAM algorithms are actively employed, particularly in autonomous vehicles and digital twin applications, due to their products of real-time sensor localization and pose, odometry, and 3D reconstruction. One of the major categories among widely applied algorithms is Visual-SLAM, which utilizes cameras, and LiDAR-SLAM, which employs LiDAR sensors. A combination of multiple sensors, known as Visual-LiDAR SLAM, has been applied in 3D reconstruction to generate the colorized point cloud of objects. It generates high-resolution images into realistic textures and colors for reconstructing real-world objects to a colorized point cloud map. LiDAR provides detailed depth and geometric information, while cameras contribute realistic textures. These can be effectively applied in precise indoor 3D reconstructions to support engineering, visualization, simulation, and many other purposes.

While many SLAM algorithms offer robustness and high precision, their implementation in indoor conditions still faces significant challenges. Indoor environments often contain fewer visual textures compared to outdoor, as common indoor objects are walls and pillars with repetitive patterns [1-3]. Rapid scene changes due to such objects positioning closer to the sensors, resulting in a parallax effect, introduce motion blur, creating blind spots and reducing overlap between frames. Additionally, the limited field of view (FoV) of the cameras in indoor spaces exacerbates these challenges. Common errors in indoor reconstruction tasks include double walls where the same structure is mapped twice with slight offsets and drifting which refers to the gradual accumulation of mismatch over time between the estimated position and orientation of sensors. These errors often go undetected during data collection, necessitating a trial-and-error approach requiring revisits and multiple repetitive scans, making SLAM costly and time-consuming.

The intrinsic challenges of current SLAM techniques lie in that, as though the real-time visualization of point cloud registration seems available, its geometric accuracy totally relies on users' perception looking into two scenes: point cloud and physical environment. This drawback can appear in a complicated layout, and the middle or at the end of the datasets. This lack of prompt feedback can delay the verification process to obtain satisfactory quality.

Mixed reality (hereafter, MR), particularly when combined with vision sensors, presents a significant opportunity to address those limitations. MR merges elements between augmented reality (AR), where digital elements, holograms, overlay the real world, and virtual reality (VR), which immerses users in a completely digital environment, enabling users to manipulate and interact with virtual objects. It blends the physical and digital worlds, enhancing user perception and interaction with spatial data for more reliable and efficient data collection and analysis. MR with RGB sensor fusion advancements not only improves geometric accuracy but also facilitates better decision-making by providing immediate feedback on the captured data. Microsoft HoloLens, a well-known MR headset, is available to realize such advanced capabilities for enhancing SLAM performance. Embedding MR visualization with SLAM technology offers immediate feedback on point cloud registration and improves the quality of the results.

To address these challenges, we developed the LiMRSF (LiDAR-MR-RGB Sensor Fusion) system, enabling users to perceive the in-situ point cloud registration by looking through an MR headset. This customized framework overlays point cloud meshes as holograms, perfectly aligned with the physical environment in real world while automatically highlighting errors detected during overlaps. The model is transmitted via a TCP server to the MR headset, where the coordinate system is calibrated to align with the world coordinates. Through the see-through glasses of the headset, users can view the localized mesh, quickly identifying blind spots and errors, and taking immediate corrective actions on-site.

The physical setup of this visual sensor fusion system integrates a high-performance LiDAR sensor and an RGB camera, with each sensor precisely fixed and calibrated, extrinsic parameters, to align color accurately to each LiDAR-generated point. This allows the generation of colorized point cloud maps with precise geometric information and realistic textures. This handheld device offers users the flexibility to move freely while capturing intricate features within indoor environments. While commercial 3D scanners also produce colorized point clouds, their manufacturer often restricts access to raw sensor data. To achieve high-fidelity colorized point clouds and mesh models, it is essential to combine raw sensor data and convert it into a transferable format with tightly integrated calibration information. Furthermore, these raw data are processed and refined to automatically detect and highlight errors, providing the user with sufficient insight to compare the generated data with the physical environment through the MR headset. The LiMRSF offers an effective solution for scanning complex geometries through real-time visualization in mixed reality, enabling the reconstruction of indoor environments. The mesh model error highlight is an important feature of this system indicating lower density point clouds than the average.

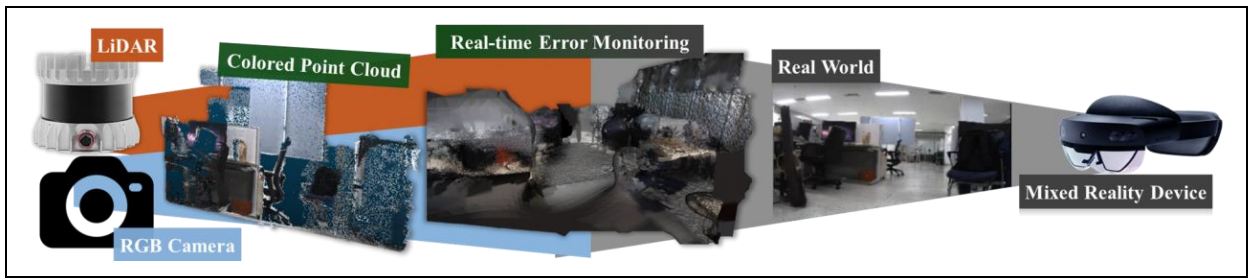

**Figure 1 LiMRSF System Overview:** a calibrated LiDAR-RGB sensor integration generates a colored point cloud and is processed as a mesh model. The MR headset visualizes the mesh 3D reconstruction model with highlighted errors overlayed on the physical environment

The growing demand for remote sensing technologies is driven by the need to reduce costs and the limitations of human labor in tasks that require extensive environmental observation. Advanced remote sensing techniques now combine multi-sensor inputs (e.g., LiDAR and cameras) with real-time processing capabilities to provide accurate, timely information for applications such as environmental monitoring, urban planning, and 3D reconstruction. Our LiMRSF system enables users to automatically monitor point cloud meshes with errors highlighted, allowing them to intuitively identify blind spots and compare the results with the physical environment to enhance the quality of 3D reconstructed models. This minimizes both the cost and labor associated with revisiting sites and the workload of manual error detection in 3D reconstruction.

The paper will include the following sections: a Literature Review of related research, a System Overview of "LiMRSF" with detailed steps of necessary calculations and software, Experimental Validation to confirm the process's effectiveness, a Conclusion, and References.

## 2. Literature Review

The convergence of SLAM, 3D Reconstruction, and Mixed Reality (MR) technologies has significantly advanced the field of indoor environment mapping and visualization. Integrating LiDAR sensors and RGB camera systems with MR devices has facilitated instantaneous visualization and interaction with reconstructed environments, which is particularly relevant in fields such as architecture, engineering, and construction (AEC). This literature review explores the foundational technologies and methodologies that underpin this study, focusing on the development and application of SLAM techniques, the role of 3D reconstruction in enhancing Building Information Modeling (BIM), and the evolution of MR technologies in creating immersive and accurate digital representations of physical spaces.

### 2.1 Simultaneous Localization and Mapping (SLAM)

Simultaneous Localization and Mapping (SLAM) is a crucial area of research in computer vision and robotics, enabling autonomous systems to map environments while tracking their location. Key approaches include Visual-SLAM, which uses camera data for detailed textures [4,5], and LiDAR-SLAM, which provides precise geometric and depth information [6]. Fusing these methods, known as Visual-LiDAR SLAM, is valuable for generating colorized point clouds in complex indoor environments. However, challenges such as limited sensor field of view, rapidly changing scenes, and close object proximity often lead to errors like double walls and drifting, requiring additional scanning sessions and increasing costs [7-9]. Recent advancements in SLAM include the development of ORB-SLAM, a monocular SLAM system recognized for its accuracy and robustness across various environments [10]. Additionally, systems like LVI-SAM, which integrate LiDAR, visual, and inertial data, have further enhanced mapping accuracy and robustness [11]. The integration of semantic information through approaches like SemanticFusion has also contributed to improved scene understanding, further enhancing SLAM's applicability in complex environments [12]. Another critical aspect of SLAM is sensor fusion and calibration. Effective integration of data from multiple sensors, such as LiDAR and RGB cameras, is essential for generating accurate and high-quality point clouds. Precise extrinsic calibration, where the relative positioning and orientation between sensors are determined, is crucial for ensuring the alignment of the color and depth information. This calibration is particularly challenging in dynamic indoor environments, where sensor data must be continuously synchronized [13].

### 2.2 3D Reconstruction and Build Information Modeling (BIM)

3D reconstruction plays a pivotal role in the digital representation of real-world objects and environments, particularly in the architecture, engineering, and construction (AEC) sectors. Accurate digital models are essential for structural analysis, project planning, and visualization. In SLAM, 3D reconstruction typically involves the generation of colorized point clouds, which combine geometric data from LiDAR sensors with texture information from RGB cameras [14]. These point clouds serve as the foundation for Building Information Modeling (BIM), a process that integrates comprehensive data about a building's geometry, materials, and systems into a single digital model. BIM models facilitate enhanced project management, structural analysis, and real-time stakeholder collaboration. The integration of 3D reconstruction with BIM is particularly advantageous in indoor environments characterized by intricate structural details. However, traditional methods of generating BIM from laser-scanned point clouds are often labor-intensive and prone to inaccuracies, especially in environments where blind spots and overlapping structures are prevalent. Recent advancements in technology have focused on automating the reconstruction of as-built BIMs from laser-scanned data, thereby improving both accuracy and efficiency [15,16]. Beyond BIM, the high-resolution 3D models generated through SLAM and 3D reconstruction have applications in various fields, such as virtual reality (VR) simulations, structural health monitoring, and cultural heritage preservation. These models are not only useful for construction but also for ongoing maintenance and inspection tasks, where precise and up-to-date digital representations are required [17,18]. The use of LiDAR and RGB data in creating detailed BIM models also supports various applications, such as construction progress monitoring, quality control, and facility management [19,20]. The integration of IoT with BIM further enhances the capabilities of smart building management, providing real-time data for efficient operations [21]. Moreover, expanding BIM from 3D to n-D modeling integrates various data types, including time, cost, and sustainability information, making it a versatile tool for managing complex construction projects [22]. The expansion of BIM into Industry 4.0 applications, such as integrating with Computational Fluid Dynamics (CFD) for enhanced simulation and analysis, exemplifies the transformative potential of this technology [23].

**2.3 Mixed Reality (MR)**

Mixed Reality (MR) represents a significant advancement in immersive technology, blending the physical and digital worlds to allow real-time interaction with both. MR builds upon Augmented Reality (AR) by ensuring that virtual objects adhere to the physical laws of the real world, enhancing the realism and utility of the interaction. In the context of SLAM and 3D reconstruction, MR devices such as see-through glasses provide a powerful platform for visualizing and interacting with reconstructed environments [24,25]. The deployment of MR devices enables real-time visualization of SLAM-generated point clouds, allowing for immediate detection and correction of errors such as drifting and double walls. This capability is particularly beneficial in indoor environments where rapid changes in the scene and limited sensor FoV often lead to incomplete or inaccurate reconstructions. By providing real-time feedback, MR technology reduces the need for repeated scanning sessions, ensuring higher quality and more accurate outputs [26,27]. Furthermore, the integration of MR with BIM enhances the ability to simulate and interact with digital models of structures, providing engineers and architects with an immersive and informative experience. This integration is particularly relevant in the context of Industry 4.0, where MR is used to enhance the efficiency and accuracy of processes such as inspection, simulation, and training [28]. As MR technology continues to evolve, it is expected to play an increasingly critical role in structural engineering and related fields, where the demand for precise and efficient visualization tools is growing due to demographic shifts and a shrinking skilled labor force [29,30].

## 3. System Overview

This section provides a detailed description of the system overview developed for applying the instantaneous visualization of the point cloud mesh model in mixed reality. The system overview is presented in Figure 3. With this framework, we aim to exploit the use of computer vision techniques, and mixed reality to localize the visualized scanned information on the see-through glasses of the MR headset in the LiMRSF device designed in this work. To generate a precise colorized point cloud map, sensors including LiDAR and camera must be tightly calibrated in millimeter precision. Python code computes the normal vectors of generated colorized points, performs Poisson Reconstruction to create a mesh model, and detects and highlights the blind spots in the mesh. A TCP endpoint connects to Unity to transmit the highlighted mesh model. In Unity, the transferred mesh model is treated as an interactable object that can be translated and rotated to localize the model precisely. As a result, once the user scans the indoor environment, the LiMRSF system visualizes a mesh model, and the user can detect the errors that occur in the indoor SLAM process.

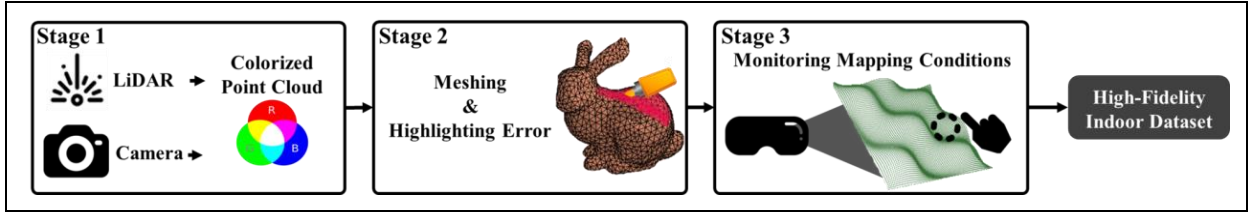

**Figure 3 System Overview;** Stage 1 creates a colorized point cloud from a tightly coupled LiDAR-RGB sensor of the LiMRSF system, Stage 2 contains meshing and highlighting errors from the scanning process, and Stage 3 a user with the LiMRSF device's see-through glasses monitors the mapping conditions to ensure the generation of high-fidelity indoor dataset.

### 3.1 Stage 1: Data Collection & Sensor Fusion

In this stage, pre-data collection is needed to generate a colorized point cloud. The sensors including the camera and LiDAR should be tightly integrated to provide a precise point cloud map as shown in Figure 3.1.1. Robot Operating System (ROS) packages are used for intrinsic and extrinsic calibration and for generating colorized point clouds.

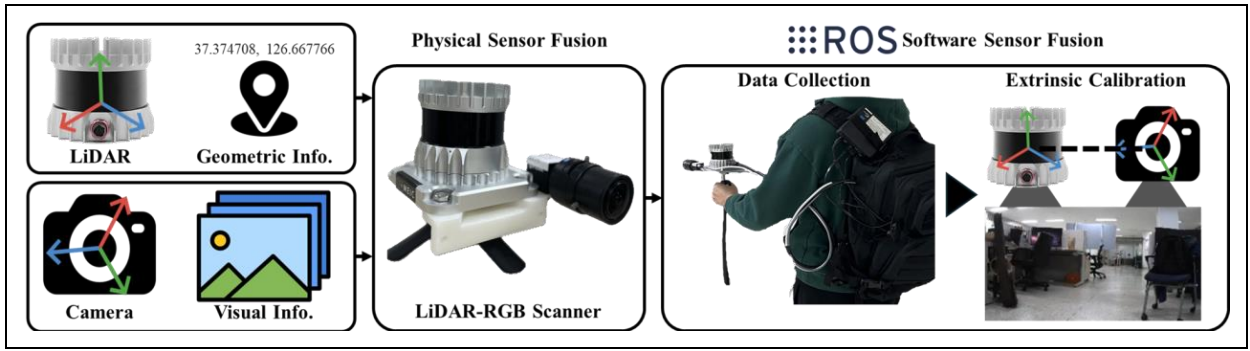

**Figure 3.1.1 Stage 1: Custom LiDAR-RGB Sensor in the LiMRSF System Overview;** LiDAR generates a point cloud that has only geometric information and the camera provides visual information and textures. The two sensors are fused in both hardware and software to generate a high-quality colorized point cloud map.

The LiDAR-RGB sensor of the LiMRSF system is built before the pre-data collection to maintain a fixed position on the sensor mount. Designed as a handheld scanner, it enables detailed scanning and minimizes blind spots. In the scanner, LiDAR generates a non-colored point cloud containing geometric information including x, y, and z coordinates. At the same time, the camera captures visual information in the form of RGB values for each pixel. Intrinsic camera calibration calculates the intrinsic parameters including focal lengths ($f_x, f_y$), principal points ($c_x, c_y$), and distortion coefficients. Extrinsic calibration calculates the extrinsic parameters, specifically the rotation matrix $R$ and translation vector $t$ [31].

As illustrated in Figure 3.1.2, an image from the camera and a point cloud map from the LiDAR are generated during pre-data collection. The extrinsic parameters are calculated based on the camera's intrinsic parameters, 2D image coordinates from camera images, and 3D world point coordinates from the LiDAR-generated point cloud maps.

Once the calibration yields a reliable rotation matrix and translation vector, the color mapping step initiates. The ROS package for LiDAR-inertial-visual state estimation and mapping uses the calibration data from the pre-data collection step [13]. The package collects sensor data from the tightly coupled LiDAR and camera, coloring the point cloud within the camera's field of view. These colorized point clouds are continuously stacked over the previous scans to create a large-scale dataset, or colorized point cloud maps.

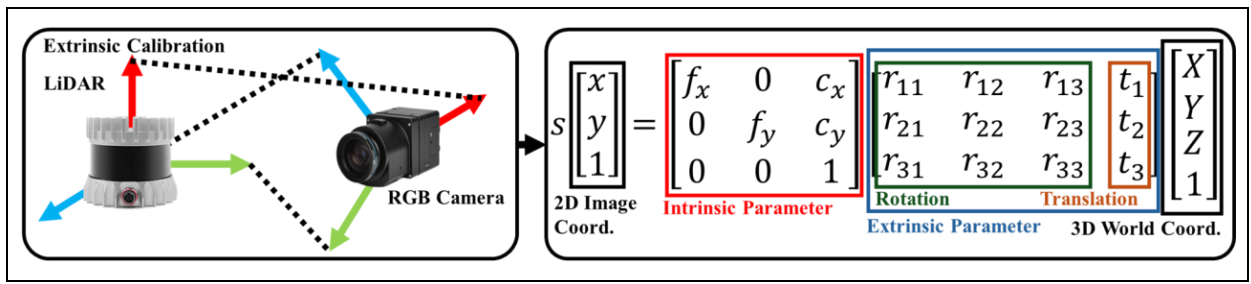

**Figure 3.1.2 Sensor Calibration Calculations;** Extrinsic calibration is finding the locations of both sensors and aligning their coordinate systems. LiDAR provides 3D world coordinates, and the camera provides 2D image coordinates. The intrinsic parameter data can be obtained from intrinsic camera calibration calculations.

## 3.2 Stage 2: Mesh and Highlighting Errors

In the initial stages, ROS utilizes the LiMRSF system to create a colorized point cloud map of the indoor environment. However, transmitting thousands of points through a wireless connection in the LiMRSF system can be challenging due to the large data size, and the point cloud's sparsity may impact the visualization in the scanned map. Rather than transmitting each point, the colorized point cloud map can be reconstructed as a mesh model. This 3D representation is created by converting a set of discrete points into a surface made up of connected polygons, smoothing surfaces like walls and ceilings. This approach significantly reduces the data size compared to the point cloud map. Moreover, to help users identify errors such as blind spots, blank areas in the mesh must be highlighted. This process involves outlier removal, normal estimation, Poisson surface reconstruction to generate the mesh, and identifying and highlighting blind spots for user detection.

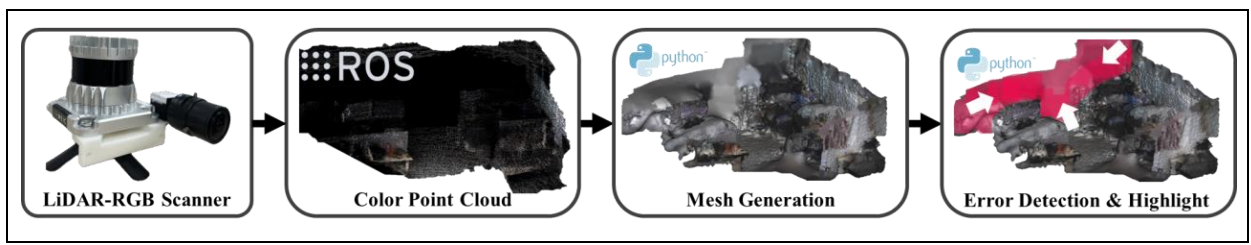

**Figure 3.2 Mesh and Highlighting Error Overview;** the LiMRSF system creates a colored point cloud. Python code generates mesh from it, detects errors, and highlights them as red for the user.

To begin, the colored point cloud from the LiMRSF system contains outliers caused by the sensor noise, reflections, and environmental factors such as lighting. These outliers must be removed to create a smooth mesh. For each point, $p_i$ in the point cloud $P = \{p_1, p_2, \ldots, p_n\}$, the average distance $d_i$ to its $k$-nearest neighbors are computed as:

$$d_i = \frac{1}{k} \sum_{j=1}^{k} dist(p_i, p_j) \qquad (1)$$

where $dist(p_i, p_j)$ is the Euclidean distance between point $p_i$ and its neighbor $p_j$.

The overall mean $\mu_d$ and standard deviation $\sigma_d$ of these distances are calculated. Points with distances exceeding a threshold are classified as outliers to be removed:

$$Threshold = \mu_d + (std_{ratio}) \times \sigma_d \tag{2}$$

Normal vectors for each point in the point cloud are computed by fitting a local plane to each point and its neighboring points as preprocessing for Poisson surface reconstruction. For each point $p_i$, the covariance matrix $C_i$ from its neighbors is calculated:

$$C_i = \frac{1}{k}\sum_{j=1}^{k}(p_j - \bar{p_i})(p_j - \bar{p_i})^T \tag{3}$$

where $\bar{p_i}$ is the centroid of the neighboring points. The eigenvector corresponding to the smallest eigenvalue of $C_i$ provides the normal vector to the local surface at the point $p_i$.

Poisson surface reconstruction is to reconstruct a smooth surface from the point cloud, utilizing the normal vectors of each point [32]. The point cloud contains both positions $(x, y, z)$ and normal vectors that describe surface orientation. It solves the Poisson equation:

$$\nabla \cdot N = \rho \tag{4}$$

where $N$ is the vector field of the normal vectors, and $\rho$ is the indicator function representing the underlying surface. The solution to equation (4) yields a surface $S$ that approximates the point cloud by fitting to the normal vectors, balancing fidelity to input data with smoothness.

The Poisson reconstruction computes density values for each vertex, representing how well the surface is supported by nearby points in the point cloud. For each vertex $v_i$, a density $d_i$ is computed based on the point density from the Poisson reconstruction. The mean density $\bar{d}$ is:

$$\bar{d} = \frac{1}{N}\sum_{i=1}^{N} d_i \tag{5}$$

Vertices with densities below a threshold are identified as blind spots and highlighted. The threshold is set as a fraction of the mean density:

$$Threshold = \mu_{density} \times density_{threshold} \tag{6}$$

where $\mu_{density}$ is the mean density value across all vertices.

Lastly, to enable fast data transfer to the MR device, the number of triangles in the mesh is reduced. The quadric error metric evaluates the cost of collapsing two vertices into one. The error for a vertex $v$ is defined as:

$$Q(v) = v^T A v + 2b^T + c \tag{7}$$

where $A$, $b$, and $c$ represent quadric matrix derived from the surrounding triangles. Simplifying the mesh minimizes this quadric error by collapsing edges and reducing the number of triangles.

### 3.3 Stage 3: Mixed Reality Visualization and Localization

In stage 3, the highlighted mesh is transferred and visualized on the see-through glasses of the MR headset in the LiMRSF device via a TCP-Endpoint connection. The internal camera of the MR headset detects the user's hand and its motion, allowing free interaction with the virtual objects, such as scaling, rotating, and translating. The transferred mesh model derived from the colorized point cloud is configured as a grabbable object, enabling the user to position it accurately over real-world structures. The TCP-Endpoint method organizes and integrates the data in Unity for smooth transmission.

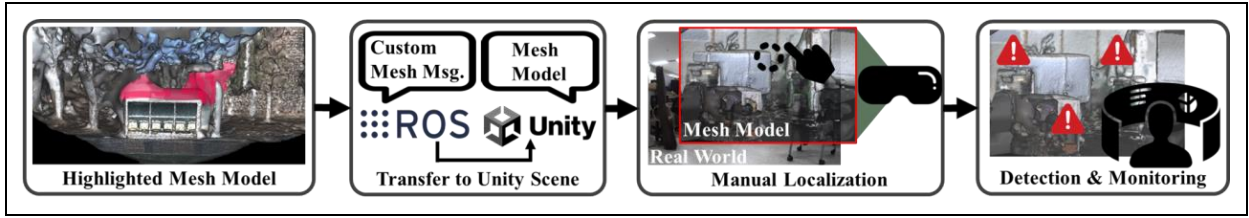

**Figure 3.3 Mixed Reality Visualization and Localization Overview; The** highlighted mesh model is transferred to Unity through TCP-server from ROS and the user localizes the received mesh model into the real world manually to detect and monitor the results.

### 3.3.1 TCP-Endpoint and Visualization

The colorized point cloud map and its mesh, highlighting any detected errors are processed on the ROS machine for efficient data management. Limited surface data is then transmitted to Unity for MR device visualization. Given the distance between the MR headset and the ROS machine, wireless transfer of the mesh model is required. For this study, ROS-TCP-Endpoint and ROS-TCP-Connector from Unity Robotics Hub are employed [33,34]. The TCP-Endpoint facilitates communication between Unity and ROS nodes over TCP/IP, commonly used for integrating Unity-based robotics simulations with ROS control and sensing systems. A custom ROS message was created, containing the Vertices, Triangles, and Vertex Colors, to handle the mesh data and prepare it for visualization on the LiMRSF system. Here, Vertices define the 3D positions, Triangles specify vertex connections to form the triangular mesh faces, and Vertex Colors allow custom coloring or transparency for rendering. This custom ROS message is translated into Unity for accepting and rendering the colored mesh from ROS.

The Unity project in the LiMRSF system contains ROS-TCP-Connector asset, MRTK, and colored mesh visualization asset. After finalizing Unity project settings and the build process without errors, the project is deployed as a .appx bundle file for the MR device. This application functions similarly to the Unity project on the PC but provides users with the mobility to explore the indoor environment without being tethered to stationary PCs.

### 3.3.2 Localization

To align the transferred mesh model with the real world, manual localization is performed by the user, offering an efficient and intuitive way to interact with the virtual model. In Unity, MRTK2 (Mixed Reality Toolkit) provides *ObjectManipulator* and *NearInteractionGrabble* features, enabling users to rotate and translate the mesh model manually. HoloLens 2, the chosen MR headset, uses advanced sensors, including time-of-flight depth sensors and cameras, to track the position of the user's hands in real time. Hand tracking involves capturing key points on the hands, known as hand joints, and mapping them into the virtual environment. HoloLens 2 calculates the 3D positions of various hand joints from sensor data and represents them as vectors in 3D space, with each joint orientation represented by quaternions to capture orientation and rotation.

The *ObjectManipulator* component allows an object to be moved or rotated by the user. Manipulation involves translating hand movements into changes in the object's transformation using rigid body transformation and spatial mapping. When the user grabs the object, the relative movement of the hand is mapped to the object's position, which is updated based on the user's hand movement:

$$P_{object\ new} = P_{object\ old} + (P_{hand\ current} + P_{hand\ start}) \tag{8}$$

where $P$ represents the positions, and $P_{hand\ start}$ is the position where the hand initially grabbed the object.

Rotation is computed by tracking the orientation of the hand as it changes during manipulation. Quaternions are used to avoid singularities and interpolation issues. If $Q_{hand\ start}$ and $Q_{hand\ current}$ are the quaternion representations of the hand's orientation at the start and current times, the object's new rotation is computed as:

$$Q_{object\ new} = Q_{hand\ current} \times Q_{hand\ start}^{-1} \times Q_{object\ old} \tag{9}$$

where $\times$ denotes quaternion multiplication.

The *NearInteractionGrabble* component enables users to grab objects through direct, physical hand interactions when they are near the object. This component does not directly contain the transformation logic but acts as a trigger for interaction based on the proximity of hand joints to the object. When a hand's palm or fingertip comes within a certain range of the object (defined as a bounding box or collider), the system considers the object "grabbable." The distance for this interaction can be computed as:

$$Distance = \sqrt{(x_{obejct} - x_{hand})^2 + (y_{obejct} - y_{hand})^2 + (z_{obejct} - z_{hand})^2} \tag{10}$$

where $x$, $y$, and $z$ are coordinates of the object and hand joints in 3D space.

### 3.4 Model Evaluation Method

In this step, the accuracy of the blind spot detector and the similarity between the generated mesh models and simplified models compared to the physical world are evaluated. First, the accuracy of the blind spot is assessed by employing different methods and calculating Precision, Recall, F1 score, and Intersection over Union (IoU) [35, 36]. These evaluation metrics quantify the number of True Positive (TP), False Positive (FP), and False Negative (FN) in the identified blind spots. The similarity evaluation measures how closely the generated mesh models that the user is monitoring resemble the physical world. This similarity assessment is performed using the Structure Similarity Index Measure (SSIM), Peak Signal-to-Noise Ratio (PSNR), and Mean Square Error (MSE) by comparing the mesh models with the ground truth images [37, 38].

Blind spot detection refers to the portions of the point cloud that were missed during scanning, which affects the meshing or modeling of the high-fidelity models and necessitates rescanning. Consequently, the detected blind spots tend to be larger than expected. This implies that small spaces (e.g., under shelves, and couches) that can be smoothed during meshing are neglected, whereas larger spaces (e.g., behind doors, and unscanned areas in the point cloud map) are recognized as blind spots.

The methods for identifying and evaluating blind spots within point cloud data involve four key stages: Point Cloud Density Estimation, Identification of low-density Regions, Mapping of Mesh Blind Spots to the Point Cloud, and the Computation of Evaluation Metrics.

Point Cloud Density Estimation quantifies the spatial distribution of points within the cloud. For each point $p_i$ in the point cloud $P$, the local density $\rho(p_i)$ is estimated by counting the number of neighboring points within a specific radius $r$:

$$\rho(p_i) = \sum_{p_j \in P} \mathbb{I}(\|p_j - p_i\| \leq r) \tag{11}$$

where $\mathbb{I}$ is the indicator function that equals 1 if the condition is true and 0 otherwise. This density measure provides a basis for distinguishing between densely populated regions and potential blind spots.

Subsequently, the Identification of low-density Regions is conducted by establishing a density threshold $T$ based on a predefined percentile $\boldsymbol{P}$ of the density distribution:

$$T = Percentile(\{\rho(p_i)\}_{i=1}^N, \boldsymbol{P} \tag{12}$$

Points with densities below this threshold ($\rho(p_i) < T$) are classified as low-density regions, which are considered blind spots. This thresholding ensures that only statistically significant low-density areas are selected for further analysis.

The next step involves Mapping Mesh Blind Spots to the Point Cloud. Mesh blind spot vertices are associated with corresponding points in the point cloud by identifying all point cloud points within a mapping radius $r_m$ of each blind spot vertex $m_j$ in the mesh $M$:

$$M = \bigcup_{m_j \in M_{blind}} \{p_i \in P | \|m_j - p_i\| \leq r_m\} \tag{13}$$

where $M_{blind}$ denotes the set of blind spot vertices in the mesh. The resulting set $M$ comprises the mapped point cloud indices corresponding to the mesh blind spots

To evaluate the accuracy of blind spot identification, the methodology computes Evaluation Metrics based on the overlap between the mapped indices $M$ and the colorized point cloud $G$. The metrics include True Positives (TP), False Positives (FP), False Negatives (FN), Precision ($P$), Recall ($R$), F1 Score ($F1$), and Intersection over Union ($IoU$). TP represents the number of correctly identified blind spot points, FP denotes points incorrectly identified as blind spots, and FN indicates colorized point cloud blind spot points that were not identified. These are defined as follows:

$$TP = |M \cap G|, \quad FP = |M - G|, \quad FN = |G - M| \tag{14}$$

Based on these, the evaluation metrics are calculated using the following equations:

$$P = \frac{TP}{TP + FP}, \quad R = \frac{TP}{TP + FN}, \quad F1 = 2 \times \frac{P \times R}{P + R}, \quad IoU = \frac{TP}{TP + FP + FN} \tag{15}$$

These metrics collectively assess the accuracy and completeness of the blind spot identification process, with higher values indicating better performance.

The similarity evaluation of the mesh model is essential for assessing how the user monitors the hologram model through the see-through glasses of the LiMRSF device. The Structure Similarity Index Measure (SSIM), Peak Signal-to-Noise Ratio (PSNR), and Mean Square Error (MSE) are employed to compare by importing ground truth images and comparing them with both the mesh models' (i.e. original mesh model, and simplified mesh model) images captured as close as the position of the ground truth images.

SSIM measures the perceptual similarity between two images, considering luminance, contrast, and structure information. For images $I_1$ and $I_2$, SSIM is defined as:

$$SSIM(I_1, I_2) = \frac{(2\mu_1\mu_2 + C_1)(2\sigma_{12} + C_2)}{(\mu_1^2 + \mu_2^2 + C_1)(\sigma_1^2 + \sigma_2^2 + C_2)} \tag{16}$$

where $\mu_1$, $\mu_2$ are the means of $I_1$ and $I_2$, $\sigma_1^2$, $\sigma_2^2$ are their variances, $\sigma_{12}$ is the covariance, and $C_1$, $C_2$ are constants to stabilize the division.

PSNR evaluates the ratio between the maximum possible power of a signal and the power of corrupting noise, providing a measure of reconstruction quality. It is calculated as:

$$PSNR = 10 \log_{10}\left(\frac{MAX^2}{MSE}\right) \tag{17}$$

where $MAX$ is the maximum possible pixel value of the image, and $MSE$ is the mean squared error between the two images $I_1$ and $I_2$ over $N$ pixels:

$$MSE(I_1, I_2) = \frac{1}{N}\sum_{i=1}^{N}\bigl(I_1(i) - I_2(i)\bigr)^2 \qquad (18)$$

where $N$ denotes the total number of pixels. Therefore, MSE quantifies the average squared difference between the corresponding pixels of the two images.

In summary, this methodology systematically estimates point cloud density, identifies low-density blind spots, maps these spots to the mesh structure, and evaluates the identification accuracy using well-defined metrics. Parameter optimization further refines the process, ensuring its effectiveness and adaptability to different datasets and conditions.

## 4. Experimental Validation

In this section, experimental validation for generating a colorized point cloud map and mesh model, which is then visualized on the LiMRSF device. This validation includes selecting an experimental environment, creating the LiDAR-RGB sensor, and setting up the LiMRSF system visualization.

### 4.1. Experimental Setup

For the experimental setup, an appropriately sized indoor environment with complicated objects — such as desks, interior objects, and pillars — is essential. The MEIC laboratory in the State University of New York, Korea, Incheon (SUNY Korea) has an ideal setting to scan, localize, and visualize the mesh. Validation will focus on the main hall, encompassing three sides of the walls. As shown in Figure 4.1, the laboratory spans approximately 120 m² and contains desks, pillars, interior objects, a robot arm, and various equipment.

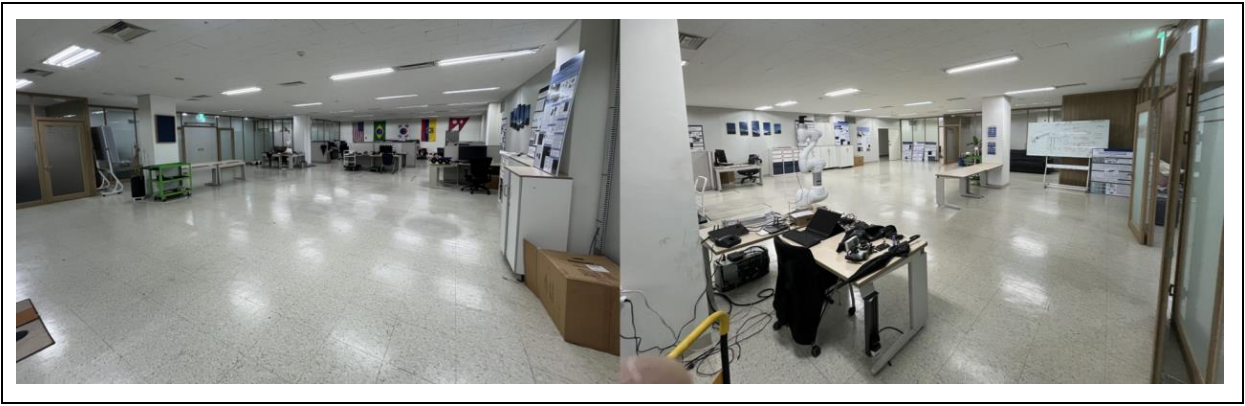

**Figure 4.1 Images of MEIC Lab. State University of New York, Korea, Incheon**

### 4.2. Equipment Setup

As detailed in Table 4.2, our system setup includes two computers, a FLIR Blackfly S for RGB imaging, an Ouster OS0-32 for LiDAR, and a Microsoft HoloLens 2 for the LiMRSF MR visualization. The Intel NUC 11[th] mobile PC serves as the ROS machine for convenient movement during indoor scanning, CPU and RAM handling the generation of colorized point cloud maps and mesh models for transfer to Unity (HoloLens 2 app). An MSI GT63 Titan 8RF laptop is used to build the Unity project and deploy the app to HoloLens 2. The LiDAR and the RGB camera connect to the ROS machine to create a colorized point cloud. HoloLens 2 runs the Unity app that connects to the ROS machine via TCP-Endpoint to visualize the mesh model on its see-through glasses.

The LiMRSF device comprises several physical components, including the LiDAR-RGB sensor and the mixed reality device, HoloLens 2, as depicted in Figure 4.2. HoloLens 2 is connected wirelessly to the ROS machine, while the LiDAR-RGB sensor, monitor, and 12V battery, along with their data transmission wires and power lines, are stored inside the backpack. The data transmission wires connect via Ethernet cables to a switching hub, which directly transmits data to the ROS machine for launching sensor drivers and receiving raw sensor data (e.g., images, and point clouds). The small monitor is attached for visualizing the colorized point cloud and monitoring the status of sensor data recording.

**Table 4.2 System Specifications**

| Intel NUC 11th (ROS Machine) | | MSI GT63 Titan 8RF (Unity Development) | |
|---|---|---|---|
| Processor | Intel 11th i7-1165G7 2.80GHz x8 | Processor | Intel 8th i7-8850H 2.60GHz x6 |
| Memory | G.Skill DDR4-3200 8GB x2 | GPU | NVIDIA GeForce GTX 1070 |
| Disk | 500GB SSD | Memory | DDR4-2667 8GB x2 |
| OS | Ubuntu 20.04.6 LTS | Disk | 500GB SSD |
| **FLIR Blackfly S (RGB Camera)** | | OS | Windows 10 Home |
| Resolution | 1440 x 1080 | **Microsoft HoloLens 2** | |
| Frame Rate | 30 FPS | Processor | Qualcomm Snapdragon 850 |
| Field of View | 70° | HPU | 2nd Holographic Process Unit |
| Camera Sensor | SONY IMX273 | Resolution | 2k 3:2 |
| CMOS Sensor Size | 9 ½ ' | Memory | 4 GB LPDDR4x DRAM |
| **Ouster OS0-32 (LiDAR)** | | Disk | 64GB UFS 2.1 |
| Horizontal Scan | 360° | Sensors | 4 RGB Cameras, 2 IR Cameras, 1 Depth Camera, IMU |
| Vertical Scan | 90° (±45°) | | |
| Max. Range | 35 m | Environmental Understanding | 6 DoF Tracking, Spatial Mapping, MR Capture |
| Resolution | 1024 x 20 Hz | | |
| Data Transfer | 655,360 points/s | Battery | 2-3 Hours of Active Use |

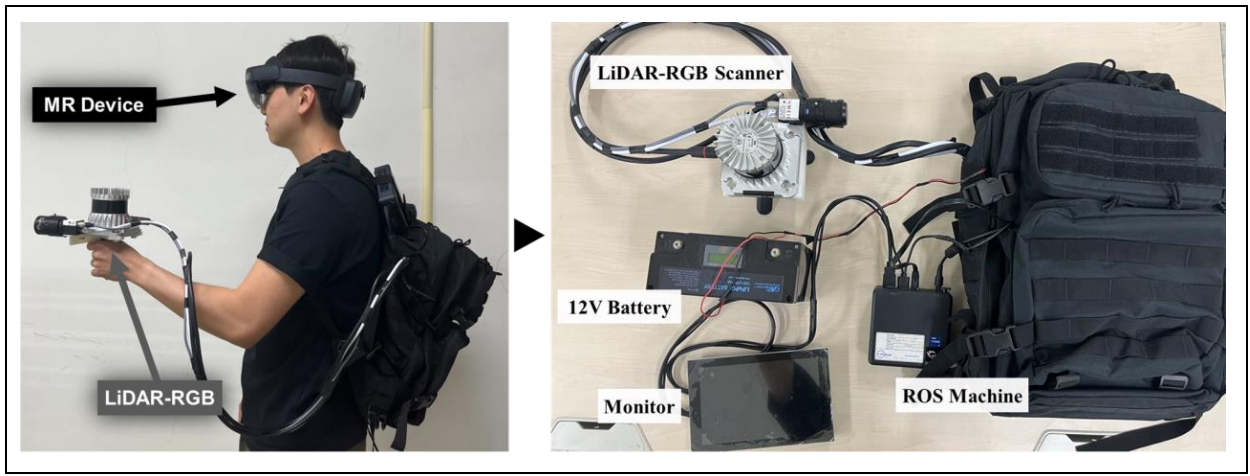

**Figure 4.2 LiMRSF Device and its Detail View;** The left image shows the user holding the LiDAR-RGB scanner and wearing the HoloLens 2, while the right image displays the LiMRSF Device components: the LiDAR-RGB scanner, ROS machine, 8-inch monitor, and 12V battery with wiring inside the backpack.

## 4.3. Error Highlighted Mesh Monitoring

The Unity app comprises three main Unity assets: *ROS-TCP-Connector*, *MeshReceiver*, and *Mixed Reality Toolkit* (hereafter, *MRTK*). *ROS-TCP-Endpoint* and *Connector*, provided by Unity Robotics Hub, allows connection with ROS machine, allowing Unity to publish or subscribe to custom ROS topics and visualize them in real-time. The *MeshReceiver* asset subscribes to the mesh model from ROS and dynamically renders it in the Unity scene, allowing users to interact with and manipulate the mesh. Lastly, *MRTK*, a Microsoft-driven project, provides essential components and features to accelerate cross-platform MR app development in Unity. In this study, we applied the *Unity OpenXR Plugin*, utilizing features like the input system, hand interaction, and spatial awareness.

For development, we used *Unity Editor 2020.3.42f1 LTS* and *Visual Studio 2019* to deploy the application on HoloLens 2. After installing the necessary packages and refining the code, the Unity project is built through the Unity Editor's build process, outputting a *.sin* solution file. This file can be debugged and built as an application for HoloLens 2 in *Visual Studio*. We used *Visual Studio 2019*'s publishing tool to create an *.appx* bundle containing the project and its dependencies. This app bundle file is then downloaded and installed through the *Windows Device Portal*, with the Unity development laptop connected to the HoloLens 2 over Wi-Fi. Once installation is complete, the app can be launched on HoloLens 2.

This workflow is divided into three stages: **Colorize Point Cloud and Generate Mesh**, **Highlight Error**, and **Error Monitoring**.

**Colorize Point Cloud and Generate Mesh**

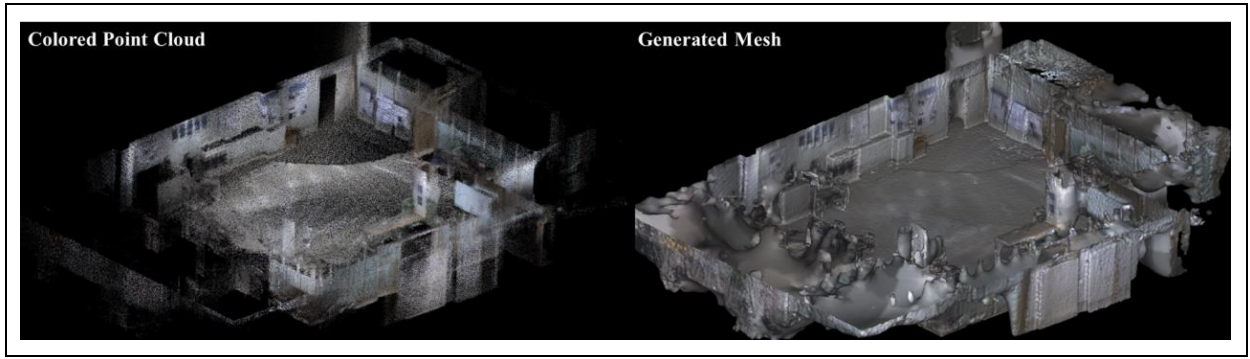

**Figure 4.3.1 Generated Colorized Point Cloud, and Mesh Providing a High-quality Virtual Model with Details of the MEIC Laboratory;** image captured from CloudCompare software and cropped the ceiling portion for visibility.

After the extrinsic calibration, precise colorization of the point cloud is achieved using the $R^3$LiVE package from GitHub, which combines LiDAR-Inertial Odometry (LIO) and Visual-Inertial Odometry (VIO) [10]. These methods construct the geometric structure of the global map through LiDAR scans while rendering its texture by minimizing frame-to-map photometric error. Parameters are adjustable depending on the usage conditions.

The colorized point cloud is then used as an input for the meshing process in Python, which involves outlier removal, normal estimation, and Poisson surface reconstruction. For outlier removal, 20 nearest neighbors are used in statistical outlier removal, with a standard deviation threshold of 2.0 to filter out points that deviate significantly from their neighbors. In normal estimation, a radius of 0.5 defines the size of the local neighborhood. For Poisson reconstruction, an octree depth of 13 is used to generate a detailed mesh. Figure 4.3.1 provides an overview of the colored point cloud and mesh generation process.

**Highlight Error & Monitoring**

In the process of generating a mesh model, blind spots are detected by comparing the point cloud density, and these sparse areas are highlighted in red on the mesh. Sparse or missing sections of point cloud data are identified when densities fall below a threshold of 0.3. These detected sparse areas are highlighted in red, allowing users to easily recognize blind spots, as shown in Figure 4.3.2.

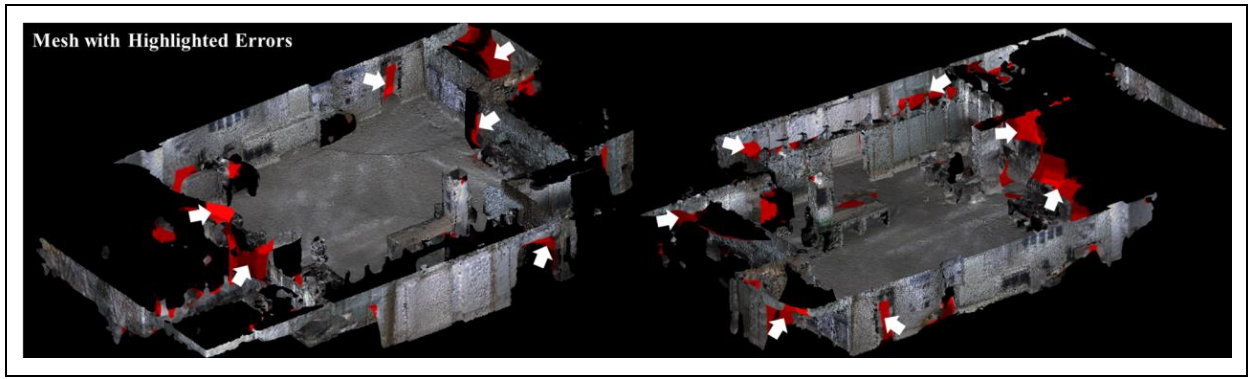

**Figure 4.3.2 Two Views of Mesh with Highlighted Detected Errors: Before Simplification;** the image captured from CloudCompare Software and cropped the ceiling portion for visibility

This generated mesh model with highlighted errors is transmitted via the TCP-Endpoint server to the Unity app deployed on HoloLens 2. The *MeshReceiver* asset subscribes to this mesh and visualizes it on the HoloLens 2's display. Due to HoloLens 2's limited rendering capabilities, the original mesh with highlighted errors, consisting of 150,000 vertices, is reduced to 10,000 vertices using Equation (7), as shown in Figure 4.3.3. If this limit is exceeded, the Unity app may fail to load all vertices, resulting in a collapsed mesh structure. Moreover, this reduction ensures a decrease in rendering delay within the LiMRSF system and avoids bottlenecks in the TCP-Endpoint server.

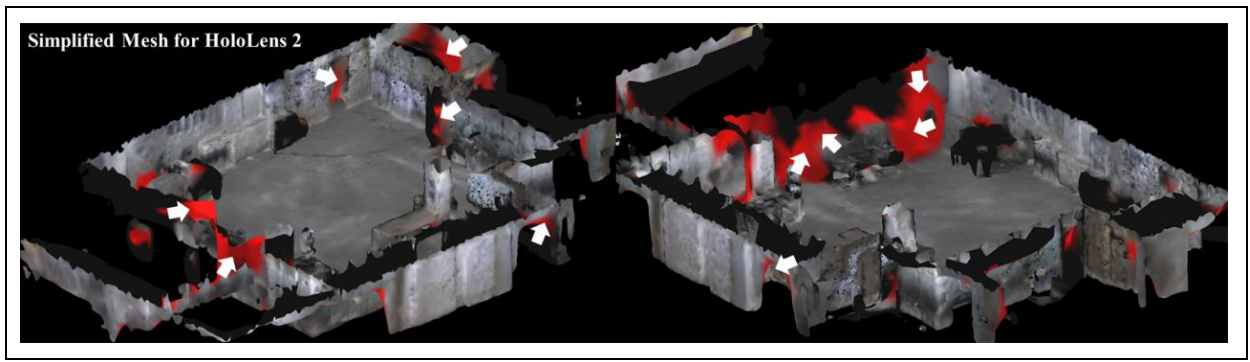

**Figure 4.3.3 Two Views of Mesh with Highlighted Detected Errors: After Simplification;** the image captured from CloudCompare Software and cropped the ceiling portion for visibility

Users can grab and position the mesh model in alignment with the structure. The mesh is scaled to real-world dimensions, as the LiMRSF system provides an accurate point cloud map. To ensure the visibility of both virtual and real objects, the entire mesh is rendered with transparent texture, while the highlighted areas are made slightly more transparent to help users identify portions of the site that may not have been scanned accurately.

### 4.4. Evaluation Metric Results

As Figure 4.3.2 illustrates, large blind spots that require rescanning are highlighted in red. These blind spots affect the entire structure of the mesh model, which is contrary to the goal of generating high-fidelity models. Consequently, these detected blind spots play a crucial role in achieving high-fidelity models, and it is imperative to ensure that the blind spot detector is accurately applied to detect the correct errors. The core of the evaluation involves calculating the number of blind spots in terms of True Positive (TP), False

Positive (FP), and False Negative (FN) by comparing the colorized point cloud with the original mesh model that has highlighted errors. In Equations (12) and (13), the percentile $P = 60$, and mapping radius $r_m = 0.5\ meters$, which is relatively large to detect significantly large blind spots for rescanning. The results of evaluating Intersection over Union (IoU), Precision, Recall, and F1 Score are presented in Table 4.4.1.

**Table 4.4.1 Blind Spot Detector Accuracy Evaluation;** Calculated IoU, Precision, Recall, and F1 Score, based on the number of True Positive (TP), False Positive (FP), and False Negative (FN).

| Method | IoU | Precision | Recall | F1 Score |
|---|---|---|---|---|
| **Full Model** | 0.6098 | 0.6554 | 0.8976 | 0.7576 |
| | **True Positive (TP)** | **False Positive (FP)** | **False Negative (FN)** | |
| | 1333390 | 701158 | 152182 | |

The results indicate that the number of correctly identified blind spots is 1,333,390 (TP), the number of points incorrectly identified as blind spots is 701,158 (FP), and the number of blind spots that were not identified in colorized point clouds is 152,182 (FN). From Equation (15), the evaluation metrics are computed as $IoU = 0.6098$, $Precision = 0.6554$, $Recall = 0.8976$, and $F1\ Score = 0.7556$. The IoU value indicates a 60.98% overlap between predicted blind spots and the actual blind spots. The Precision value signifies that 65.54% of the identified blind spots by the original mesh with highlighted errors are true blind spots. The Recall value indicates that the original mesh model successfully identifies 89.76% of all actual blind spots. Lastly, an F1 Score of 0.7556 reflects a balanced performance between Precision and Recall. As a result, a high number of FP indicates over-predicting blind spots; however, the results demonstrate that our blind spot detector achieves 75.56% accuracy.

The mesh model with highlighted errors (i.e., the original mesh) and the simplified mesh model for HoloLens 2 visualization are used as inputs to evaluate model similarity. As described in Section 3.4, Model Evaluation Method, this step assesses whether the mesh model that users are monitoring (i.e., the simplified mesh model) is sufficiently similar to the actual physical world. We applied Equations (16), (17), and (18) to evaluate the similarities between the ground truth images, the original mesh, and the simplified mesh. The original mesh model is also evaluated to ensure that the generated mesh from the colorized point cloud accurately visualizes the structure produced by our LiDAR-RGB sensor in the LiMRSF system.

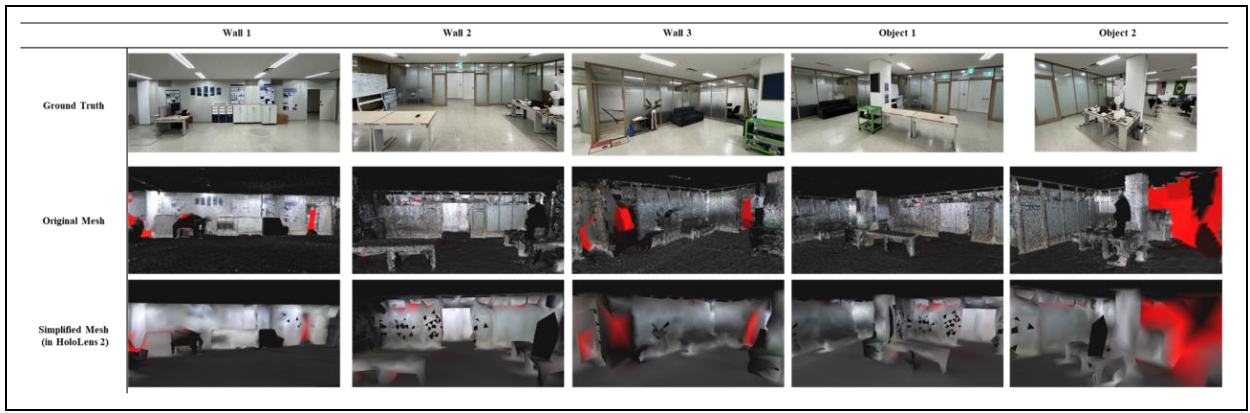

**Figure 4.4 Cropped and Captured Original and Simplified Mesh Models of the Experimental Space;** Ground truth images are captured with an iPhone 13 mini camera (0.5x zoom). The original mesh model and simplified mesh model are screen captures from CloudCompare software, captured from the locations close to where the ground truth images were taken.

To evaluate the similarities, the Structure Similarity Index Measure (SSIM), Peak Signal-to-Noise Ratio (PSNR), and Mean Square Error (MSE) methods are applied. The ground truth images and captured images of the original mesh and simplified mesh are obtained from CloudCompare software, taken from positions closely aligned with those of the ground truth image. SSIM values close to 1.0 and higher PSNR values indicate high similarity to the ground truth, while lower MSE values signify higher similarity between the images. The results are shown in Table 4.4.2.

**Table 4.4.2 Evaluated SSIM, PSNR, and MSE Values;** The values are separated as ground truth vs. original mesh model and ground truth vs. simplified mesh model across five different sections. SSIM values closer to 1.0 and higher PSNR values indicate structural similarity, while lower MSE values indicate higher similarity between the images.

|  | Ground Truth vs Original Mesh Model | | | Ground Truth vs Simplified Mesh Model | | |
| --- | --- | --- | --- | --- | --- | --- |
| Method | SSIM | PSNR | MSE | SSIM | PSNR | MSE |
| Wall 1 | 0.349 | 8.9055 | 0.1287 | 0.3942 | 9.2222 | 0.1196 |
| Wall 2 | 0.2762 | 9.987 | 0.1003 | 0.365 | 10.0141 | 0.0997 |
| Wall 3 | 0.4012 | 13.0169 | 0.0499 | 0.5619 | 14.1004 | 0.0389 |
| Object 1 | 0.3379 | 11.3702 | 0.0729 | 0.4619 | 12.3725 | 0.0579 |
| Object 2 | 0.3784 | 12.5184 | 0.056 | 0.5337 | 14.0958 | 0.0389 |

The mesh models are divided into five different sections: three walls and two objects. The Wall 2 section exhibits higher SSIM and PSNR values along with the lowest MSE values, indicating that this section has the highest similarity to the ground truth. However, most of the SSIM values are not close enough, averaging around 50%, and PSNR values are also significantly low. Low SSIM values suggest noticeable structural or perceptual differences between ground truth and the mesh models, implying potential loss of fine details or structural integrity in the mesh models. Moreover, low PSNR values indicate that the mesh models differ significantly from the ground truth. In visual or perceptual terms, a low PSNR usually points to less accurate reconstructions.

Nevertheless, the combination of low SSIM and PSNR with low MSE indicates that while the reconstructions may appear different (structurally or perceptually) from the ground truth, they maintain a relatively low average error. This could be because certain areas of the models are well-preserved while others deviate structurally, leading to overall low MSE but noticeable differences in structure or texture captured by SSIM and PSNR. Moreover, as described in Equation (4), the surface $S$ that approximates the point cloud by fitting the estimated normal vectors to smooth the surfaces, resulting in the loss of detail. Additionally, as shown in Figure 4.4, the zoomed views of mesh models have rough surfaces compared to the ground truth images.

In conclusion, the inherent characteristics of Poisson Surface Reconstructed models lead to low SSIM and PSNR values, but low MSE values. Nonetheless, visual similarity is sufficient for users to monitor and detect blind spots for generating high-fidelity models.

### 4.5. Visualization Results

This section validates how the simplified mesh model is visualized as holograms in our LiMRSF device (i.e. HoloLens 2). The original mesh model, comprising 150,000 vertices, is simplified to 10,000 vertices to mitigate delays in rendering and transmission through the TCP-Endpoint server. The simplified mesh model is then transferred to the Unity application deployed on the HoloLens 2 via the server. Upon receiving the data, the Unity application displays the simplified mesh model as holograms overlaying the see-through glasses of the HoloLens 2.

Users can interact with the holographic model by pinching to adjust by rotating or translating it to position the model accurately within the physical environment. This interactive capability allows for precise alignment of the hologram with real-world structures, enhancing the user's ability to monitor and assess the spatial accuracy of the mesh model. The visualized results are illustrated in Figure 4.5.

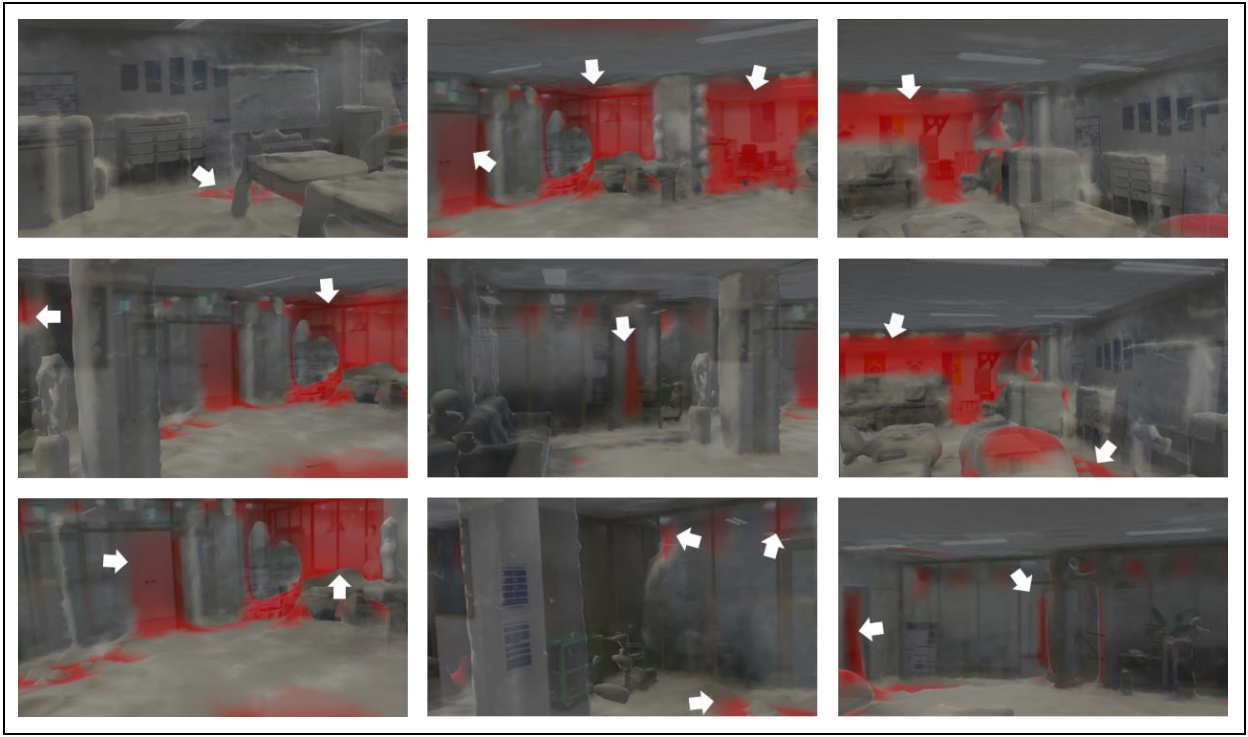

**Figure 4.5 LiMRSF Visualization in HoloLens 2;** Captured images from the recorded video in Windows Device Portal, which can access the HoloLens 2's front cameras

## 5. Conclusion

In conclusion, we have demonstrated that our system effectively detects blind spots in the 3D reconstruction process, thereby reducing the time and cost associated with site revisits for repetitive scanning. We successfully visualized the 3D reconstructed mesh model with highlighted errors on the see-through glasses of the MR device in near real-time. The tightly integrated LiDAR-RGB sensor in the LiMRSF system produces a highly accurate colorized point cloud map, enabling the creation of a detailed textured mesh. Highlighted errors allow users to identify blind spots within the scanned colorized point cloud easily. Additionally, the TCP-Endpoint server provides a reliable wireless connection for transferring the mesh model to the Unity application.

As shown in Section 4.4, we have evaluated the accuracy of our blind spot detector and the similarity between the original mesh with highlighted errors and the simplified mesh model to the ground truth images. The blind spot detector was assessed using IoU, Precision, Recall, and F1 Score metrics. Specifically, the F1 score was 0.7576, indicating that our blind spot detector can detect errors with 75.76% accuracy. For similarity evaluation, we applied SSIM, PSNR, and MSE methods. The mesh models' images were captured from locations corresponding to the ground truth images, and the metrics were computed accordingly. Although SSIM and PSNR values were significantly low, the low MSE values and the nature of the Poisson Surface Reconstruction mesh models demonstrate that these SSIM and PSNR values are acceptable. Moreover, as shown in Section 4.5, the visualization of simplified mesh models closely resembles the physical world's properties, allowing users to intuitively recognize the models.

The rendering performance of the HoloLens 2 imposes a limitation by necessitating the reduction of the mesh from 150,000 vertices to 10,000 vertices. Since HoloLens 2 hardware is over five years old, its processing power is insufficient to display high-resolution mesh models. While this constraint limits the display of a fully detailed mesh model, the simplified version is transparent enough to maintain visibility and accurate error detection.

The mesh generation and error highlighting processes take approximately three minutes, which could be expedited with a desktop computer with higher computation power. However, to maintain user mobility, the ROS machine must remain within proximity. With high-performance MR devices and mini-PCs, real-time monitoring and error detection from the SLAM-based 3D reconstruction can be achieved more seamlessly.

In conclusion, we have developed a non-existing system, the LiMRSF system, that reconstructs mesh models from point cloud registration using the LiDAR-RGB sensor and successfully visualizes them as holograms with blind spot detection on the LiMRSF device (i.e. HoloLens 2) by applying mixed reality technology. Our proposed system can indicate blind spot errors by highlighting the area on the mesh, allowing users to monitor the condition of the reconstructed model and determine the need for rescanning. A refreshed 3D reconstructed model, updated by rescanning the error sites, provides a high-fidelity colorized point cloud map, thereby enhancing Building Information Modeling (BIM).